\documentclass[aps,prx,twocolumn,superscriptaddress,amsmath,amssymb]{revtex4-2}
\usepackage{graphicx}
\usepackage{dcolumn}
\usepackage{bm}
\usepackage{color}
\usepackage{amsmath}
\usepackage{bbold}
\usepackage{braket}
\usepackage{mathtools}
\usepackage{dsfont}
\usepackage{amssymb}
\usepackage{xcolor}
\setcitestyle{super}
\usepackage{caption}
\usepackage{subcaption}

\renewcommand{\vec}[1]{{\boldsymbol #1}}

\usepackage{letltxmacro}
\LetLtxMacro{\originaleqref}{\eqref}

\usepackage{etoolbox}
\pretocmd{\eqref}{Eq.~}{}{}
\usepackage{siunitx}
\newcommand{\en}{\varepsilon}

\renewcommand{\vec}[1]{{\boldsymbol #1}}

\usepackage{dsfont}

\usepackage[justification=raggedright,font=small,margin=0pt]{caption}

\begin{document}

\title{
Giant magnetochiral anisotropy from quantum confined surface states\\
of topological insulator nanowires}

\author{Henry F. Legg}
\altaffiliation{These authors contributed equally}
\affiliation{Department of Physics, University of Basel, Klingelbergstrasse 82, CH-4056 Basel, Switzerland}

\author{Matthias R{\"o}{\ss}ler}
\altaffiliation{These authors contributed equally}
\affiliation{Physics Institute II, University of Cologne, Z{\"u}lpicher Str. 77, 50937 K{\"o}ln, Germany}

\author{Felix M{\"u}nning}
\affiliation{Physics Institute II, University of Cologne, Z{\"u}lpicher Str. 77, 50937 K{\"o}ln, Germany}

\author{Dingxun Fan}
\affiliation{Physics Institute II, University of Cologne, Z{\"u}lpicher Str. 77, 50937 K{\"o}ln, Germany}

\author{Oliver Breunig}
\affiliation{Physics Institute II, University of Cologne, Z{\"u}lpicher Str. 77, 50937 K{\"o}ln, Germany}

\author{Andrea Bliesener}
\affiliation{Physics Institute II, University of Cologne, Z{\"u}lpicher Str. 77, 50937 K{\"o}ln, Germany}

\author{Gertjan Lippertz}
\affiliation{Physics Institute II, University of Cologne, Z{\"u}lpicher Str. 77, 50937 K{\"o}ln, Germany}

\affiliation{KU Leuven, Quantum Solid State Physics, Celestijnenlaan 200 D, 3001 Leuven, Belgium}
\author{Anjana Uday}
\affiliation{Physics Institute II, University of Cologne, Z{\"u}lpicher Str. 77, 50937 K{\"o}ln, Germany}

\author{A. A. Taskin}
\affiliation{Physics Institute II, University of Cologne, Z{\"u}lpicher Str. 77, 50937 K{\"o}ln, Germany}

\author{Daniel Loss}
\affiliation{Department of Physics, University of Basel, Klingelbergstrasse 82, CH-4056 Basel, Switzerland}
\author{Jelena Klinovaja}
\affiliation{Department of Physics, University of Basel, Klingelbergstrasse 82, CH-4056 Basel, Switzerland}
\author{Yoichi Ando}
\affiliation{Physics Institute II, University of Cologne, Z{\"u}lpicher Str. 77, 50937 K{\"o}ln, Germany}

\maketitle


{\bf Wireless technology relies on the conversion of alternating electromagnetic fields to direct currents, a process known as rectification. While rectifiers are normally based on semiconductor diodes, quantum mechanical non-reciprocal transport effects that enable highly controllable rectification have recently been discovered\cite{Ando2020, Isobe2020, Yasuda2016, Yasuda2020, Baumgartner2021, Rikken2001, Rikken2005, Tokura2018, He2018}. One such effect is magnetochiral anisotropy (MCA)\cite{Rikken2001, Rikken2005, Tokura2018, He2018}, where the resistance of a material or a device depends on both the direction of current flow and an applied magnetic field. 
However, the size of rectification possible due to MCA is usually extremely small, because MCA relies on inversion symmetry breaking leading to the manifestation of spin-orbit coupling, which is a relativistic effect\cite{Rikken2001, Rikken2005, Tokura2018}. In typical materials the rectification coefficient $\gamma$ due to MCA is usually\cite{Tokura2018, Morimoto2016, Ideue2017, He2018, Rikken2019} $|\gamma| \lesssim 1$ ${\rm A^{-1} T^{-1}}$ and the maximum values reported so far are $|\gamma| \sim 100$ ${\rm A^{-1} T^{-1}}$ in carbon nanotubes\cite{Krstic2002} and ZrTe$_5$\cite{Wang2020}. Here, to overcome this limitation, we 
artificially break inversion symmetry via an applied gate voltage in thin topological insulator (TI) nanowire heterostructures and theoretically predict that such a symmetry breaking can lead to a giant MCA effect. Our prediction is confirmed via experiments on thin bulk-insulating (Bi$_{1-x}$Sb$_{x}$)$_2$Te$_3$ TI nanowires, in which we observe an MCA consistent with theory and  $|\gamma| \sim 100000$ ${\rm A^{-1} T^{-1}}$, the largest ever reported  MCA rectification coefficient in a normal conductor.}

In most materials transport is well described by Ohm's law, $V=I R_0$, dictating that for small currents $I$ the voltage drop across a material is proportional to a constant resistance $R_0$. Junctions that explicitly break inversion symmetry, for instance semiconductor $pn$-junctions, can produce a difference in resistance $R$ as a current flows in one or the opposite direction through the junction, $R(+I)\neq R(-I)$; this difference in resistance is the key ingredient required to build a rectifier. A much greater degree of control over the rectification effect can be achieved when a similar non-reciprocity of resistance exists as a property of a material rather than a junction. However, to achieve such a non-reciprocity necessitates that the inversion symmetry of the material is itself broken. Previously, large non-reciprocal effects were observed in materials where inversion symmetry breaking resulted in strong spin-orbit coupling (SOC)\cite{Rikken2001, Rikken2005, Tokura2018, He2018, Morimoto2016, Ideue2017, Rikken2019, Wang2020}. However, since SOC is always a very small energy scale, this limits the possible size of any rectification effect. 

\begin{figure*}[t]
\includegraphics[width=0.8\textwidth]{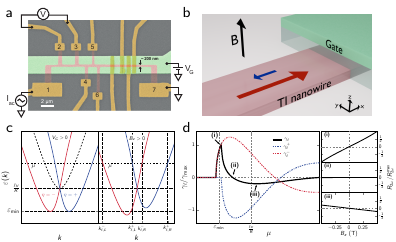}
\caption{\linespread{1.05}\selectfont{}
{\bf Gate-tunable topological insulator nanowire device and the theory of magnetochiral anisotropy: } 
\textbf{a,} False-colour scanning-electron-microscope image of Device 1 with schematics of the electrical wiring; the Pt/Au leads are coloured in dark yellow, the TI nanowire etched from an MBE-grown BST thin film in red, and the top-gate electrode in green. The resistance of the nanowire was measured on different sections: Section 1, 2, 3, 4, 5 correspond to the voltage-contact pairs 2-3, 3-4, 4-5, 5-6, and 2-6, respectively. 
\textbf{b,} Schematic of MCA in TI nanowires. A gate, applied here to the top of the nanowire, breaks inversion symmetry along the wire. Applying a magnetic field along the gate normal ($z$-direction) results in a giant MCA rectification such that current flows more easily in one direction along the wire than the opposite (indicated by red/blue arrows).
\textbf{c,} TI nanowire surface states form degenerate subbands (dashed line). When a finite gate voltage is applied, inversion symmetry is broken and the subbands split (solid lines). A new minimum occurs at $\varepsilon_{\rm min}$ and the states possess a finite spin polarisation in the $yz$-plane (red/blue colour). A magnetic field $\vec B$ shifts the subband pair relative to each other in energy due to the Zeeman effect, which is maximal for $\vec B$ along the $z$-axis, leading to an MCA (the size of the shift shown here is not to scale and used for clarity). 
\textbf{d,} Size of the MCA rectification $\gamma_\ell$ (see Eq.~(1)) as a function of chemical potential $\mu$ within a given subband pair. Due to the peculiar dispersion of a TI nanowire, the curvature, $\hbar^2 \mathcal{V}_{\ell}^{\eta}(k) \equiv \partial^2_k \en^{\eta}_{\ell}(k)$, is large and highly anisotropic at opposite Fermi momenta resulting in a giant MCA. As the chemical potential $\mu$ is tuned from the bottom of the subband, $\gamma_\ell$ changes sign.  Here, for clarity, we used $B$ = 1 T, see Supplementary Note 5 for further parameters. The panels \textbf{e-g} show the theoretically expected magnetic-field dependence of the 2nd harmonic resistance $R_{2\omega}$ at the chemical potentials  indicated in the main panel.}\label{fig:Fig1}
\end{figure*}

The non-reciprocal transport effect considered here is magnetochiral anisotropy (MCA), which occurs when both inversion and time-reversal symmetry are broken\cite{Rikken2001, Rikken2005, Tokura2018, He2018, Morimoto2016, Ideue2017, Rikken2019, Wang2020}. When allowed, the leading order correction of  Ohm's law due to MCA is a term second order in current and manifests itself as a resistance of the form $R=R_0(1+\gamma B I)$, with $B$ the magnitude of an external magnetic field and where $\gamma$ determines the size of the possible rectification effect. MCA may also be called bilinear magnetoelectric resistance\cite{He2018, Zhang2018}. We note non-reciprocal transport in ferromagnets\cite{Yasuda2016, Yasuda2020} does not allow the coefficient $\gamma$ to be calculated and rectification of light into dc current due to bulk photovoltaic effects\cite{Inglot2015, Zhang2019, Bhalla2020} concerns much higher energy scales than MCA. 

\begin{figure*}[t]
\includegraphics[width=0.7\textwidth]{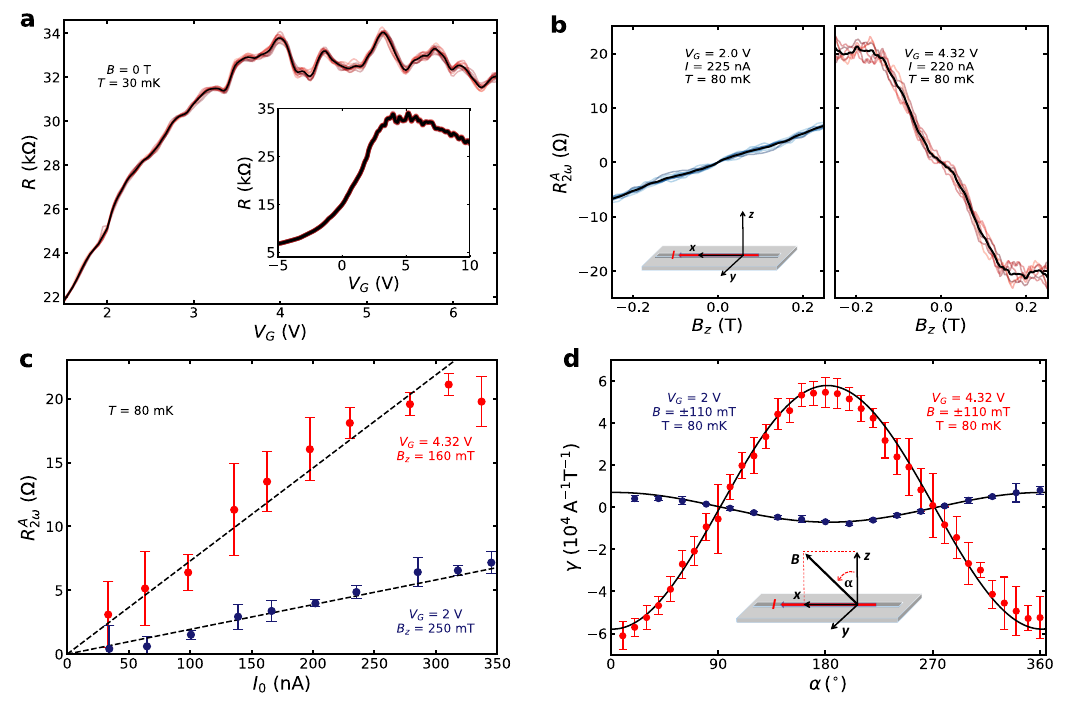}
\caption{\linespread{1.05}\selectfont{}
{\bf Non-reciprocal transport in (Bi$_{1-x}$Sb$_{x}$)$_2$Te$_3$ topological insulator nanowire:} \textbf{a,} Four-terminal resistance $R$ measured on Device 1, Section 1, at 30 mK in 0 T as a function of gate voltage $V_G$ showing reproducible peaks and dips around the resistance maximum, which are consistent with the response expected from quantum-confined surface states\cite{Munning2021}. Since the $R$ value is very sensitive to the details of the charge distributions in/near the nanowires, the $R(V_G)$ behaviour is slightly different for different sweeps; thin red lines show the results of 15 uni-directional $V_G$ sweeps and the thick black line shows their mean average. Inset shows the data for a wider range of $V_G$, demonstrating the typical behaviour of a bulk-insulating TI. 
\textbf{b,} Antisymmetric component of the second-harmonic resistance, $R^A_{2\omega}$, for $V_G$ = 2 and 4.32 V plotted vs magnetic field $\vec B$ applied along the $z$-direction (coordinate system is depicted in the inset); coloured thin lines show 10 (6) individual $B$-field sweeps for 2 V (4.32 V) and the thick black line shows their mean.
\textbf{c,} $R^A_{2\omega}$ measured for $V_{G}$ = 2 and 4.32 V in the $B$-field (applied in the $z$-direction) of 0.25 and 0.16 T, respectively, as a function of the ac excitation current $I_0$. The dashed lines are a guide to the eye marking the linear behaviour. Error bars are defined using the standard deviation of 10 (6) individual $B$-field sweeps for 2 V (4.32 V).
\textbf{d,} Magnetic-field-orientation dependencies of $\gamma$ at $V_{G}$ = 2 V (blue) and 4.32 V (red) when the $B$-field is rotated in the $zx$-plane. Error bars are defined using the Min-Max method with 6 (8) individual $B$-field sweeps for 2 V (4.32 V).. Solid black lines are fits to $\gamma \approx \gamma_0 \cos\alpha$ expected for MCA. The inset shows the definition of $\alpha$ and the coordinate system. }
\label{fig:Fig2}
\end{figure*}

In heterostructures of topological materials it is possible to artificially break the inversion symmetry of a material\cite{Legg2021}; such an approach provides an unexplored playground to significantly enhance the size of non-reciprocal transport effects.  In this context, quasi one-dimensional (1D) bulk-insulating three-dimensional TI nanowires\cite{Zhang2010, Ziegler2018, deJuan2019, Munning2021, Legg2021} are the perfect platform to investigate large possible MCA due to artificial inversion symmetry breaking. 
In the absence of symmetry breaking, for an idealised cylindrical TI nanowire -- although generalisable to an arbitrary cross-section\cite{deJuan2019,Legg2021} -- the surface states form energy subbands of momentum $k$ along the nanowire and half-integer angular momentum $\ell=\pm\frac{1}{2},\frac{3}{2},\dots$ around the nanowire, where the half-integer values are due to spin-momentum locking. The presence of inversion symmetry along a TI nanowire requires that the subbands with angular momenta $\pm \ell$ are degenerate. It is possible to artificially break the inversion symmetry along the wire, for instance, by application of a gate-voltage from the top of the TI nanowire\cite{Ziegler2018,Munning2021,Legg2021}. Such a gate voltage induces a non-uniformity of charge density across the nanowire cross-section which breaks the subband degeneracy and results in a splitting of the subband at finite momenta\cite{Legg2021} (see Fig.~\ref{fig:Fig1}c). An additional consequence is that the subband states develop finite spin polarisation in the plane perpendicular to the nanowire axis (i.e. $yz$-plane) with the states with opposite momenta being polarized in the opposite directions such that the time-reversal symmetry is respected.  When a magnetic field is applied, the subbands can be shifted in energy via the Zeeman effect, which suggests that an MCA can be present in this setup. Indeed, using the Boltzmann equation\cite{Morimoto2016,Ideue2017,Wang2020} (see Supplementary Note 4), we find an MCA of the vector product type $\gamma \propto \vec P\cdot(\hat{\vec B} \times \hat{\vec I})$ with the characteristic vector $\vec P$ in the $yz$ plane.  For the rectification effect $\gamma_\ell(\mu)$ of a given subband pair $\eta=\pm$ labelled by $\ell>0$, we find 
\begin{equation}
\gamma_\ell=\gamma^+_\ell+\gamma_\ell^-\approx \frac{e^3}{(\sigma^{(1)})^2 h B} \sum_{\eta=\pm} \tau^2 \left[\mathcal{V} _{\ell}^{\eta}(k^\eta_{\ell,R})-\mathcal{V} _{\ell}^{\eta}(k^\eta_{\ell,L})\right],
\end{equation}
where $e$ is the elementary charge, $h$ is the Planck constant, $\sigma^{(1)}$ is the conductivity in linear response, $\tau$ is the scattering time, $\mathcal{V}_{\ell}^{\eta}(k)=\frac{1}{\hbar^2}\partial^2_k \en^{\eta}_{\ell}(k)$ with $\en^{\eta}_{\ell}(k)$ describing the energy spectrum in the presence of symmetry breaking terms and of the finite magnetic field $B$ (see Fig.~\ref{fig:Fig1}c and Supplementary Note 4), and $k^\eta_{\ell,R(L)}$ is the right (left) Fermi momentum of a given subband  (see Fig.~\ref{fig:Fig1}c). Due to the non-parabolic spectrum of subbands, the difference in $\mathcal{V}_{\ell}^{\eta}(k)$ is large for a TI nanowire resulting in the giant MCA. The quantities $\gamma^+_\ell$ and $\gamma_\ell^-$ are the contributions of the individual subbands. The behaviour of $\gamma_\ell$ as a function of chemical potential $\mu$ is shown in Fig.~\ref{fig:Fig1}d.  We find that, as the chemical potential is tuned through the subband pair, $\gamma_\ell$  will change sign depending on the chemical potential. This makes the rectification effect due to the MCA highly controllable by both magnetic field direction and by the chemical potential $\mu$ within a given subband pair, which can be experimentally adjusted by a small change in gate voltage. For reasonable experimental parameters we predict that the theoretical size of the rectification can easily reach giant values $\gamma \sim 5\times10^5$~${\rm T^{-1}A^{-1}}$ (see Supplementary Note 5).

To experimentally investigate the predicted non-reciprocal transport behaviour, we fabricated nanowire devices\cite{Breunig2021} of the bulk-insulating TI material (Bi$_{1-x}$Sb$_{x}$)$_2$Te$_3$ as shown in Fig.~1a by etching high-quality thin films grown by molecular beam epitaxy (MBE). The nanowires have a rectangular cross-section of height $h\approx 16$ nm and width $w\approx$ 200 nm, with channel lengths up to several $\mu$m. The long channel lengths suppress coherent transport effects such as universal conductance fluctuations and the cross-sectional perimeter allows for the formation of well-defined subbands (see Supplementary Note 8). An electrostatic gate electrode is placed on top of the transport channel for the dual purpose of breaking inversion symmetry and tuning the chemical potential. The resistance $R$ of the nanowire shows a broad maximum as a function of the gate voltage $V_G$ (see Fig. 2a inset), which indicates that the chemical potential can be tuned across the charge neutrality point (CNP) of the surface-state Dirac cone; the dominant surface transport in these nanowires is further documented in the Supplementary Note 7. Near the broad maximum (i.e. around the CNP), the $V_G$ dependence of $R$ shows reproducible peaks and dips (see Fig.~2a), which is a manifestation of the quantum-confined quasi-1D subbands realized in TI nanowires\cite{Munning2021} --- each peak corresponds to the crossing of a subband minima, although the feature can be smeared by disorder\cite{Munning2021}. 
To measure the non-reciprocal transport, we used a low-frequency ac excitation current $I=I_0 \sin \omega t$ and probed the second-harmonic resistance $R_{2\omega}$. The MCA causes a second-harmonic signal that is antisymmetric with magnetic field $\vec B$, and therefore we calculated the antisymmetric component $R^A_{2\omega} \equiv \frac{R_{2\omega}(\vec B)-R_{2\omega}(-\vec B)}{2}$, which is proportional to $\gamma$ via $R^A_{2\omega} = \frac{1}{2} \gamma R_0 B I_0 \approx \frac{1}{2} \gamma R B I_0$, where $R_0$ is the reciprocal resistance (see Methods for details).

\begin{figure*}[t]
\includegraphics[width=0.7\textwidth]{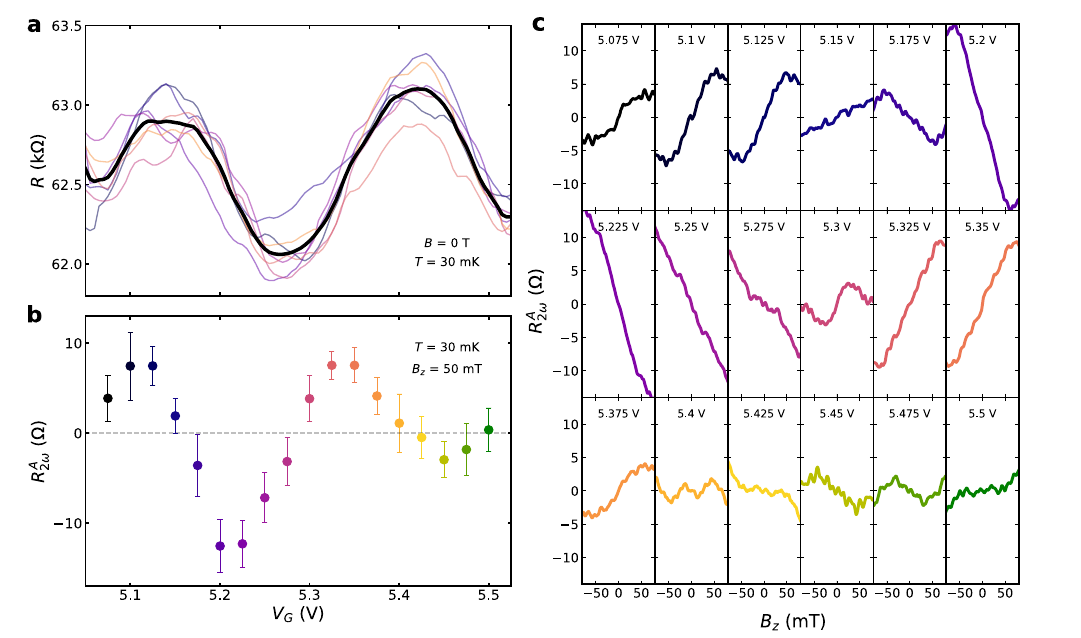}
\caption{\linespread{1.05}\selectfont{}
{\bf Evolution of the non-reciprocal response with changing chemical potential:} 
\textbf{a,} $R$ vs $V_{G}$ data of Device 3, Section 5, in a narrow range of $V_{G}$, in which the chemical potential is changed near the CNP (for a wider range of $V_G$, see Supplementary Fig.~S11a). The peaks in $R$ occur when the bottom of one of the quantum-confined subbands is crossed by the chemical potential; coloured thin lines show 7 individual $V_G$ sweeps and the thick black line shows their mean.  
\textbf{b,} Mean $R^A_{2\omega}$ values at $B_z$ = 50 mT for various gate voltages in the range corresponding to panel \textbf{a}. The zero-crossings of $R^A_{2\omega}$ roughly correspond to the peaks and dips in $R(V_G)$, and thereby are linked to the quantum-confined subbands. Error bars are defined using the standard deviation of 10 individual $B$-field sweeps.
\textbf{c,} Averaged $R^A_{2\omega}(B_z)$ curves at various $V_{G}$ settings, from which the data points in panel \textbf{b} were calculated (data points and curves are coloured correspondingly). The systematic change in the $R^A_{2\omega}(B_z)$ behaviour as a function of gate voltage is clearly visible.}
\label{fig:Fig3}
\end{figure*}

In our experiment, we observed a large $R^A_{2\omega}$ for $V_G \gtrsim$ 2 V with a magnetic field along the $z$ axis. The $R^A_{2\omega}(B_z)$ behaviour is linear for small $B_z$ (see Fig.~2b) and $R^A_{2\omega}$ increases linearly with $I_0$ up to $\sim$250 nA (see Fig. 2c), both of which are the defining characteristics of the MCA. The deviation from $B$-linear behaviour at higher fields is likely due to orbital effects (see Supplementary Note 3). The magnetic-field-orientation dependence of $\gamma$, shown in Fig. 2d for the $zx$-plane rotation, agrees well with $\gamma \approx \gamma_0\cos\alpha$, with $\alpha$ the angle from the $z$-direction and $\gamma_0$ the value at $\alpha=0$; the $yz$-plane rotation gave similar results, while MCA remained essentially zero for the $xy$-plane rotation (see Supplementary Note 10). This points to the vector-product type MCA, $R^A_{2\omega} \propto \vec P \cdot (\vec B \times \vec I)$, with the characteristic vector $\vec P$ essentially parallel to $y$, which is likely dictated by the large $g$-factor anisotropy\cite{Liu2010} (see Supplementary Note 2). The maximum size of the $|\gamma|$ in Fig.~2d reaches a giant value of $|\gamma|\sim 6\times10^4$ A$^{-1}$T$^{-1}$. 
In addition, one may notice in Figs.~2b and 2d that the relative sign of $\gamma$ changes for different $V_G$ values, which is very unusual. We observe a giant MCA with similarly large rectification $\gamma$ in all measured devices, some of them reaching $\sim 1\times10^5$ A$^{-1}$T$^{-1}$ (see Supplementary Note 13). 
Note that in the MCA literature, $\gamma$ is often multiplied by the cross-sectional area $A$ of the sample to give $\gamma'$ (= $\gamma A$), which is useful for comparing MCA in different materials as a bulk property. However, in nano-devices like our TI nanowires, the large MCA owes partly to mesoscopic effects and $\gamma'$ is not very meaningful. In fact, the large MCA rectification of $|\gamma| \sim 100$ ${\rm A^{-1} T^{-1}}$ observed in chiral carbon nanotubes\cite{Krstic2002} was largely due to the fact that a nanotube can be considered a quasi 1D system. In the Supplementary Note 13, we present extensive comparisons of the nonreciprocal transport reported for various systems.

A unique feature of the predicted MCA is the controllability of its sign with a small change of $V_G$. To confirm this prediction, we measured detailed $V_G$-dependence of $R^A_{2\omega}$ in the $V_G$ range of 5.1--5.5 V, where the chemical potential appears to pass through two subband minima, because $R(V_G)$ presents two peaks (see Fig.~3a). We indeed observe the slope of $R^A_{2\omega}(B_z)$ to change sign with $V_G$ (see Fig.~3b), and its zero-crossing roughly coincides with the peak or dip in the $R(V_G)$ curve (compare Figs. 3a and 3b). A change in sign of the slope of $R^A_{2\omega}(B_z)$ on either side of $R(V_G)$ peaks is also observed in other devices (see Supplementary Note 11). To obtain confidence in this striking observation, the evolution of the $R^A_{2\omega}(B_z)$ behaviour upon changing $V_G$ is shown in Fig.~3c for many $V_G$ values. This sign change upon a small change of gate voltage also endows the giant MCA in TI nanowires with an unprecedented level of control. In addition, this $V_G$-dependent sign change of MCA gives a unique proof that the origin of the peak-and-dip feature in $R(V_G)$ is indeed subband-crossings.

The giant MCA observed here due to an artificial breaking of inversion symmetry in TI nanowires not only results in a maximum rectification coefficient $\gamma$ that is orders of magnitude larger than any previously reported, but this giant MCA is also highly controllable by small changes of chemical potential. Although rather different to the MCA of a normal conductor discussed here, we note that large rectification effects of similar magnitude have recently been discovered in non-centrosymmetric superconductor devices\cite{Ando2020, Baumgartner2021} and in quantum anomalous Hall edge states\cite{Yasuda2020}, where the controllability is comparatively limited. It is prudent to mention that the MCA reported here was measured below 0.1 K and it diminishes around 10 K (see Supplementary Note 12), which is consistent with the subband gap of $\sim$1 meV. Since TI nanowire devices are still in their infancy\cite{Breunig2021}, the magnitude and temperature dependence of the MCA could be improved with future improvements in nanowire quality and geometry; for example, in a 20-nm diameter nanowire, the subband gap would be $\sim$10 meV which enables MCA up to $\sim$100 K. The presence of the giant MCA provides compelling evidence for a large spin-splitting of the subbands in TI nanowires with broken inversion symmetry which can be used for spin-filters\cite{streda2003,braunecker2010}. Moreover, it has  been suggested that the helical spin polarization and large energy scales possible in such TI nanowires with broken inversion symmetry can be used as a platform for robust Majorana bound states\cite{Legg2021}, which are an integral building block for future topological quantum computers.

\vspace{15pt}
\begin{flushleft} 
{\bf Acknowledgements: }
We acknowledge useful discussions with A. Rosch and B. Shklovskii.
This work was supported by the Georg H. Endress Foundation [H.F.L.] and NCCR QSIT, a National Centre of Excellence in Research, funded by the Swiss National Science Foundation (grant number 51NF40-185902) [H.F.L., D.L., J.K.]. This project has received funding from the European Research Council (ERC) under the European Union's Horizon 2020 research and innovation programme (grant agreement No 741121 [Y.A.] and grant agreement No 757725 [J.K.]). It was also funded by the Deutsche Forschungsgemeinschaft (DFG, German Research Foundation) under CRC 1238 - 277146847 (Subprojects A04 and B01) [Y.A., O.B., A.T.] as well as under Germany's Excellence Strategy - Cluster of Excellence Matter and Light for Quantum Computing (ML4Q) EXC 2004/1 - 390534769 [Y.A.]. G.L. acknowledges the support by the KU Leuven BOF and Research Foundation Flanders (FWO, Belgium), file No. 27531 and No. 52751.
\end{flushleft} 

\begin{flushleft} 
{\bf Author contributions:}  H.F.L. with help from J.K., D.L., and Y.A. conceived the project. H.F.L. with help from J.K. and D.L. performed the theoretical calculations. M.R. fabricated the devices, performed the experiments and analysed the data with help from H.F.L, F.M., D.F., O.B, and Y.A. A.B., G.L, A.U. and A.T. provided the material. H.F.L., M.R., D.L., J.K., and Y.A. wrote the manuscript with inputs from all authors.
\end{flushleft} 

\begin{flushleft} 
{\bf Competing interests:} The authors declare no competing interests.
\end{flushleft} 

\begin{flushleft} 
{\bf Correspondence:}  Correspondence and requests for materials should be addressed to 
H.F.L. (henry.legg@unibas.ch),
J.K. (jelena.klinovaja@unibas.ch), or Y.A. (ando@ph2.uni-koeln.de).
\end{flushleft} 

\begin{flushleft} 
{\bf Data availability:} The data that support the findings of this study are available at the online depository figshare with the identifier doi:10.6084/m9.figshare.19336571 (Ref.~\citenum{rossler2022}) and supplementary information. Additional data are available from the corresponding authors upon reasonable request.
\end{flushleft} 

\begin{flushleft} 
{\bf Publication note:} This version of the article has been accepted for publication, after peer review, but is not the Version of Record and does not reflect post-acceptance improvements, or any corrections. The Version of Record is available online at: \href{https://www.nature.com/articles/s41565-022-01124-1}{https://www.nature.com/articles/s41565-022-01124-1}
\end{flushleft} 


\section*{Methods}

\subsection{Theory.} Transport coefficients were calculated using the Boltzmann equation\cite{Ideue2017,Wang2020} to attain the current density due to an electric field $E$ up to second order such that $j=j^{(1)}+j^{(2)}=\sigma^{(1)} E +\sigma^{(2)} E^2$. As discussed in Ref.~\citenum{Ideue2017}, experimentally the voltage drop $V=EL$ as a function of current $I$ is measured in the form $V=R_0I(1+\gamma B I)$. 
Using that $R_0=L/\sigma^{(1)}$ for a nanowire of length $L$, a comparison to the experimental behaviour can then be achieved via the relation $\gamma_0 =-\frac{\sigma^{(2)}}{B(\sigma^{(1)})^2}$. Although the linear response conductivity $\sigma^{(1)}$ contains small peaks and dips due to an increased scattering rate close to the bottom of a subband\cite{}, such fluctuations occur on top of a large constant conductivity and we therefore approximate $\gamma_0 \approx \frac{A}{B} \sigma^{(2)}$, with $A=-1/(\sigma^{(1)})^2$  approximately constant.

 \subsection{Material growth and device fabrication.}
A 2$\times$2 cm$^2$ thin film of (Bi$_{1-x}$Sb$_{x}$)$_2$Te$_3$ was grown on a sapphire (0001) substrate by co-evaporation of high-purity Bi, Sb, and Te in a ultra-high vacuum MBE chamber. The flux of Bi and Sb was optimized to obtain the most bulk-insulating films which was achieved with a ratio of 1:6.8. The thickness varied in the range of 14--19 nm in the whole film. Immediately after taking the film out of the MBE chamber, it was capped with a 3-nm-thick Al$_2$O$_3$ capping layer grown by atomic-layer deposition (ALD) at 80 ${\rm ^o C}$ using Ultratec Savannah S200. The carrier density and the mobility of the film were extracted from Hall measurements performed at 2 K using a Quantum Design PPMS.
Gate-tuneable multi-terminal nanowire devices were fabricated using the following top-down approach: After defining the nanowire pattern with electron-beam lithography, the film was first dry-etched using low-power Ar plasma and then wet-etched with H$_2$SO$_4$/H$_2$O$_2$/H$_2$O aqueous solution. To prepare contact leads, the Al$_2$O$_3$ capping layer was removed in heated Aluminum etchant (Type-D, Transene) and 5/45 nm Pt/Au contacts were deposited by UHV-sputtering. Then the whole device was capped with a 40-nm-thick Al$_2$O$_3$ dielectric grown by ALD at 80 $^{\circ}$C, after which the 5/40nm Pt/Au top gate was sputter-deposited. Scanning electron microscopy was used to determine the nanowire size. Devices 1-4 reported in this paper were fabricated on the same film in one batch, whilst device 5 (see Supplementary Notes 7 and 8) was fabricated on a similar film.

\subsection{Second-harmonic resistance measurement.}
Transport measurements were performed in a dry dilution refrigerator (Oxford Instruments TRITON 200, base temperature $\sim$20 mK) equipped with a 6/1/1-T superconducting vector magnet. The first- and second-harmonic voltages were measured in a standard four-terminal configuration with a low-frequency lock-in technique at 13.37 Hz using NF Corporation LI5645 lock-ins. In the presence of the vector-product-type MCA with $\mathbf{P}\| \hat{\vec y}$, the voltage is given by $V = R_0 I (1+\gamma BI)$ for $\mathbf{I}\| \hat{\vec x}$ and $\mathbf{B}\| \hat{\vec z}$ . For an ac current $I = I_0 \sin \omega t$, this becomes $V = R_0 I_0 \sin \omega t + \frac{1}{2} \gamma R_0 B I_0^2 [1 + \sin (2\omega t - \frac{\pi}{2})]$, allowing us to identify $R_{2\omega} = \frac{1}{2} \gamma R_0 B I_0$ by measuring the out-of-phase component of the ac voltage at the frequency of $2\omega$. The dc gate voltage was applied by using a Keithley 2450.

\subsection{Error bars.}
In the plots of $R^A_{2\omega}$ vs $I$ shown in Fig. 2c (and in Figs. S9b, S10b, S11b, and S12b), the data points for each current value are calculated by obtaining slopes from linear fits to $R^A_{2\omega}(B)$ data at that current in the indicated $B$-range (done individually for each measured $B$-sweep); the standard deviation is calculated for the set of obtained slopes at each current and used as the error bar. In the plots of $\gamma$ vs angle shown in Fig. 2d (and in Figs. S6 and S7) as well as the plot of $\gamma$ vs $T$ shown in Fig. S13, the data points for each angle are calculated by obtaining slopes from linear fits to $R^A_{2\omega}(B)$ data at that angle in the indicated $B$-range (done individually for each measured $B$-sweep); from the set of obtained slopes at each angle, the error is calculated by using a Min-Max approach, in which we calculate the error to be half of the difference between the maximum and the minimum (calculating standard deviation gives very similar results). In the plots of $R^A_{2\omega}$ vs $V_G$ shown Fig. 3b, the data points for each $V_G$ value are calculated by obtaining slopes from linear fits to $R^A_{2\omega}(B)$ data (shown in Fig. 3c) at that $V_G$ in the indicated $B$-range (done individually for each measured $B$-sweep); from the set of obtained slopes per $V_G$, the standard deviation is calculated and used as the error bar.

\end{document}